\documentclass[a4paper]{jpconf}

\usepackage{graphicx,times}
\usepackage{array}
\usepackage{bm}
\usepackage{amsfonts}

\newcommand{\beq}{\begin{equation}}
\newcommand{\eeq}{\end{equation}}
\newcommand{\beqa}{\begin{eqnarray}}
\newcommand{\eeqa}{\end{eqnarray}}

\newcommand{\chib}{{\overline{\chi}}}
\newcommand{\phib}{{\bar{\phi}}}
\newcommand{\psib}{{\overline{\psi}}}

\newcommand\comment[1]{}
\begin{document}
\title{Fermion bag solutions to some sign problems in four-fermion field theories}

\author{Anyi Li}

\address{ Institute for Nuclear Theory, University of Washington, Seattle, 98195, USA}

\ead{anyili@uw.edu}

\begin{abstract}
Lattice four-fermion models containing $N$ flavors of staggered fermions, that are invariant under $Z_2$ and $U(1)$ chiral symmetries, are known to suffer from sign problems when formulated using the auxiliary field approach. Although these problems have been ignored in previous studies, they can be severe. In this talk, we show that the sign problems disappear when the models are formulated in the fermion bag approach, allowing us to solve them rigorously for the first time.
\end{abstract}

\section{Introduction}
Four-fermion field theories are interesting in both condensed matter and particle physics.  The well known Hubbard model and its variants are often used in studying cuprate superconductors~\cite{PhysRevLett.87.257003}, antiferromagnets \cite{PhysRevLett.98.117205} and more recently graphene~\cite{Meng:2010}. Low energy nuclear physics is also studied with four-fermion couplings in the effective field theory framework \cite{Bedaque:2002mn,Epelbaum:2009pd}. In the context of more fundamental theories like QCD, four-fermion field theories offer a simpler setting to study phenomena like fermion mass generation and chiral symmetry breaking \cite{Rosenstein:1990nm}. It has been suggested recently that quantum critical phenomena in graphene can be studied with four-fermion field theories \cite{herbut:085116,Armour:2011hf}. It has also been found there is a QCD-like sign problem in the four-fermion field theory~\cite{Grabowska:2012ik}. Despite the wide interest, strongly coupled four-fermion field theories remain poorly understood as compared to their bosonic counterparts due to computational difficulties.

The only available method to compute quantities in a strongly interacting field theory with no small parameter is the Monte Carlo (MC) method. Due to the quantum nature of a fermion there are no natural fermion configurations with positive weights that can be used for important sampling. In two space-time dimensions fermions can often be bosonized and models can we written in terms of world line configurations with positive weights. This fact can be used to design powerful MC methods \cite{Wolff:2007ip,Wolff:2008xa,Wenger:2008tq}. In higher dimensions, the traditional MC approach is to integrate the fermions out in favor of a determinant of a large fermion matrix. Whenever this determinant is positive a non-local probability distribution emerges, which can be used to construct a MC method. The most popular is the Hybrid Monte Carlo (HMC) method \cite{PhysRevB.36.8632, Duane:1987de} which has continued to evolve in many ways since its discovery \cite{Clark:2006fx}. Unfortunately, small eigenvalues of the fermion matrix which naturally arise in the presence of massless fermions can cause singularities in the HMC approach. This makes it difficult to study quantum critical phenomena containing massless fermions. While other determinantal MC methods do not encounter such problems, they scale poorly with system size \cite{Assad:2008}. In cases where the determinant of the fermion matrix is not positive, the original theory is said to suffer from a sign problem and the traditional approach is not useful. The repulsive Hubbard model away from half filling is a classic example where progress has been limited due to sign problems. Other relativistic four-fermion field theories like the Gross-Neveu (GN) models and Nambu-Jona-Lasinio (NJL) models are also known to suffer from sign problems in three or more space-time dimensions \cite{AnnPhys.224.29}.

Recently a new approach called the fermion bag approach was proposed to solve some four-fermion field theories~\cite{Chandrasekharan:2009wc, Chandrasekharan:2011vy,Chandrasekharan:2011mn,Chandrasekharan:2012fk,Chandrasekharan:2012va}. It is an extension of the meron cluster idea proposed some time ago \cite{PhysRevLett.83.3116}. The idea behind the fermion bag is to identify fermion degrees of freedom that cause sign problems and collect them in a bag and sum only over them. This is in contrast to traditional approaches where all fermion degrees of freedom in the entire thermodynamic volume are summed to solve the sign problem. When the fermion bag contains only a small fraction of all the degrees of freedom and the summation can be performed quickly, the fermion bag approach can be used to design powerful MC methods. Sometimes, the bag splits into many disconnected pieces further simplifying the calculation. The fermion bag approach has three main advantages: (a) Due to a duality, fermion bag sizes are small both at weak and strong couplings, (b) Singularities in the massless limit can be tackled without a problem, (c) Some sign problems that haunt traditional approaches are naturally solved. While the first two advantages have been demonstrated, the third advantage is not so clear from previous work. Here we show how solutions to some unsolved sign problems in four-fermion models also emerge naturally in the fermion bag approach. 

It is useful to clarify some confusions that may arise about what we mean by a sign problem and thus a solution to the sign problem. If one can write the partition of a quantum statistical mechanics system as a sum over configurations whose Boltzmann weights are all positive and if the cost of computation of the Boltzmann weights only scales as a polynomial in system size, then we say the model does not suffer from a sign problem. However, as already stated above, in fermionic systems there are no natural configurations where the Boltzmann weights are positive. The conventional method is to use the auxiliary field approach to expand the partition function as a sum of bosonic configurations where the fermion determinant is taken as part of the Boltzmann weight. If this weight can be negative one often says the model suffers from a sign problem. However, if an alternate approach can be found where the sign problem disappears, one can of course say the problem never suffered from a sign problem to begin with. On the other hand, if this alternate approach was not known earlier, the new approach can be considered as a solution to the sign problem present in the other method. This is what we mean when we say ``solutions to unsolved sign problems''. It must be noted that all sign problems are problems in exactly this sense. Once a solution is found there is no longer a problem. It is of course likely that some problems may remain unsolved \cite{Troyer:2004ge}.

We consider lattice GN models containing $N$ flavors of massless staggered fermions with either a $Z_2$ or a $U(1)$ chiral symmetry \cite{AnnPhys.224.29}. While we work in three space-time dimensions, our results can easily be extended to higher dimensions. Although in three dimensions the symmetries we refer to are a part of a flavor symmetry, they are often loosely called chiral symmetries in the literature. The $Z_2$ models with odd $N$ and all the $U(1)$ models are known to suffer from a sign problem when formulated in the traditional auxiliary field approach. Here we show that the sign problems disappear in the fermion bag approach. Our paper is organized as follows. In section 2 we review the auxiliary field approach to lattice GN models with both $Z_2$ and $U(1)$ chiral symmetries and discuss how the sign problems arise. In section 3 we discuss the severity of the sign problems. In section 4 we discuss the fermion bag approach and show that sign problems do not arise. Section 5 contains our conclusions.

\section{Auxiliary field approach}
\label{afa}

Lattice GN models are formulated in the auxiliary field approach through the action 
\beq
S_{GN}  = \sum_{x,y,i}\chib_i(x) (D[\phib])_{x,y} \chi_i(y) + S_{AF}
\label{eq:model}
\eeq
where  $\chib_i(x), \chi_i(x)$ denote the Grassmann valued fermion fields of flavor $i = 1,2..,N$ at the lattice site $x$.
The explicit form of the auxiliary field action $S_{AF}$ depends on the GN model and will be discussed below. The matrix $D[\phib]$ is defined by
\beq
\left(D[\phib]\right)_{xy}  =  D_{xy}  + \delta_{xy}\ \phib(x),
\label{eq:Dirac}
\eeq
where $\phib(x)$ is a function of the auxiliary fields as defined below and $D_{x,y}$ is the free staggered fermion matrix \cite{Sharatchandra:1981si,Kogut:1974ag,Banks:1975gq},
\beq
D_{x,y} = m \delta_{x,y} + \sum_{\alpha = 1,2,3}\ \frac{\eta_{x,\alpha}}{2}\left[\delta_{x+\alpha,y} - \delta_{x,y+\alpha}\right].
\eeq
Since we work in three dimensions, $\alpha$ labels the three directions, $\eta_{x, \alpha}= e^{(i \pi \zeta_a \cdot x)}, \zeta_1=(0,0,0)$, $\zeta_2=(1,0,0)$, $\zeta_3=(1,1,0)$ are the staggered fermion phase factors and $m$ is the bare fermion mass. We assume anti-periodic boundary conditions in all directions and denote the lattice volume by $V = L^3$.

\begin{figure}[h]
\begin{center}
\includegraphics[width=1.3in]{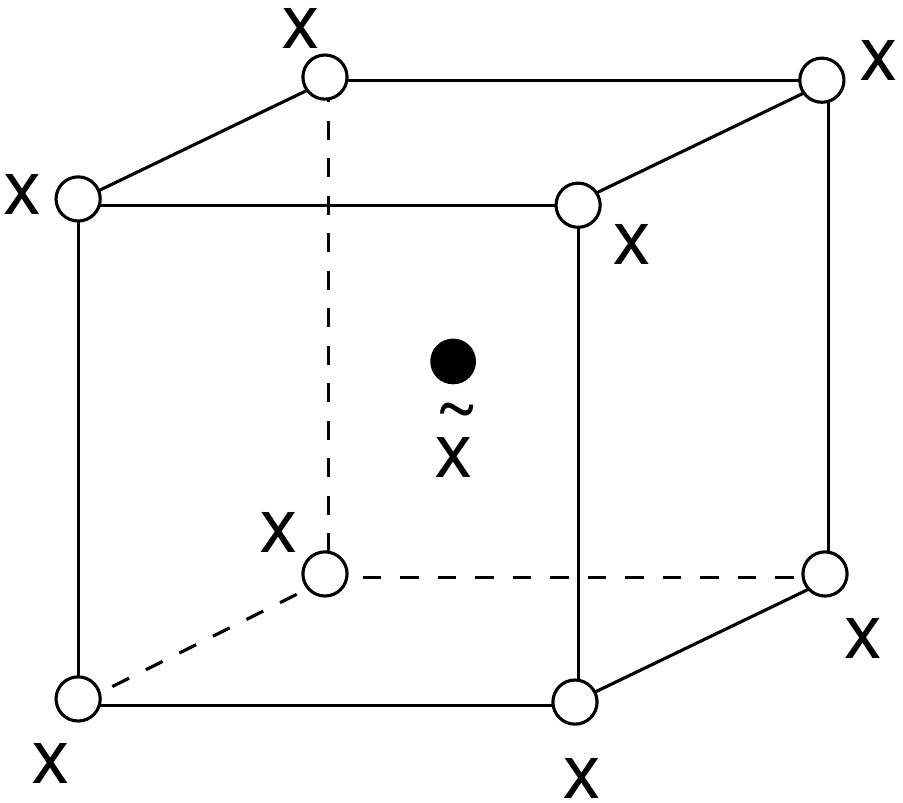}
\hskip1.0in
\includegraphics[width=1.3in]{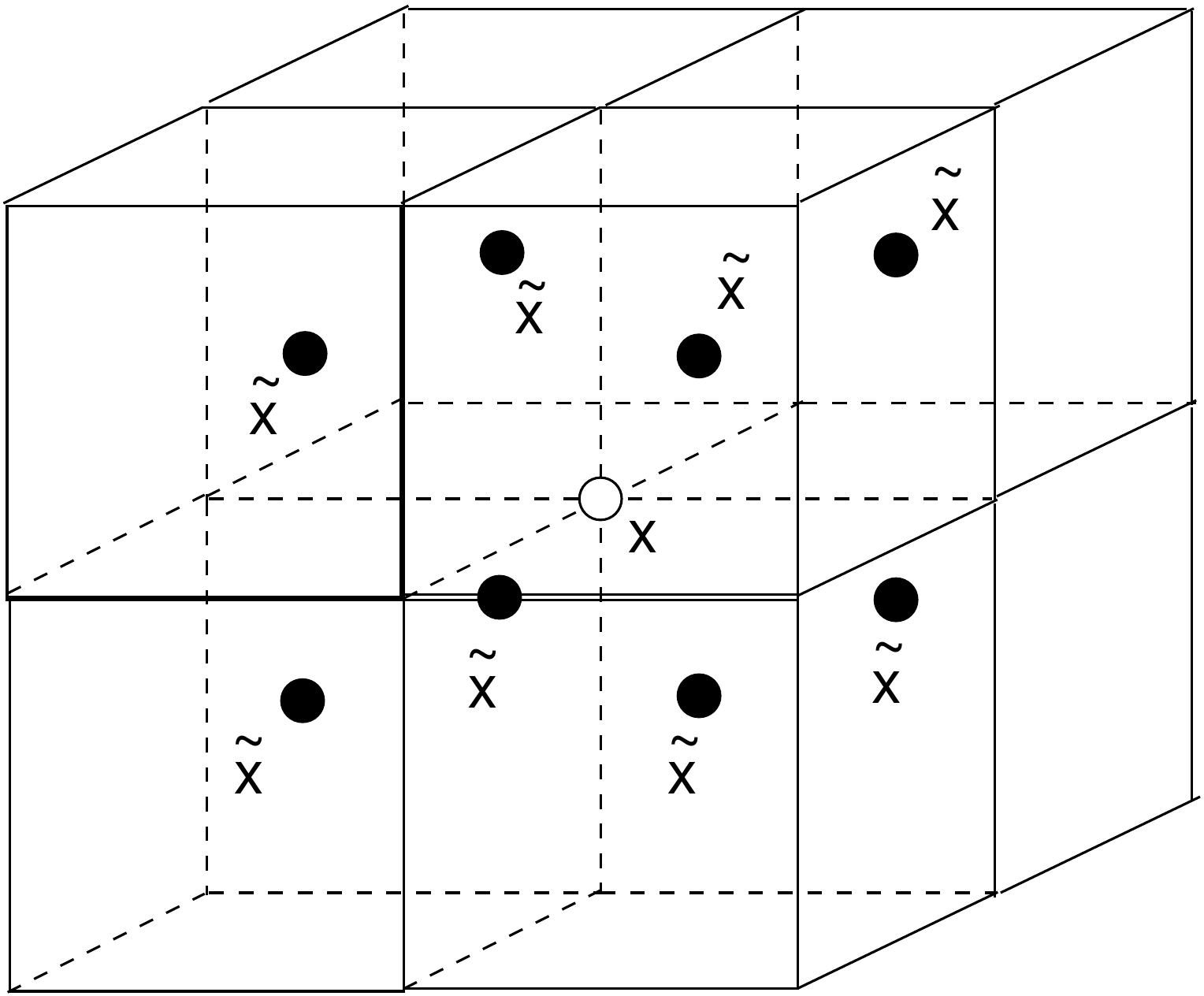}
\end{center}
\caption{Nearest neighbor lattice sites (open circles) of a fixed dual site (filled circle) $\tilde{x}$  is represented by $[ x,\tilde{x}]$ (see left figure), while the nearest neighbor dual sites of a fixed lattice site is denoted by $\langle \tilde{x},x\rangle$ (see right figure).}
\label{fig:sites}
\end{figure}

Following \cite{AnnPhys.224.29}, we define the auxiliary fields on dual sites $\tilde{x}$. The model with a $Z_2$ chiral symmetry is defined through a single real auxiliary field $\sigma(\tilde{x})$, such that 

\beqa
S_{AF}[\sigma] &=& \frac{N}{2g^2}\sum_{\tilde{x}}\ \sigma^2(\tilde{x}),
\\ 
\phib(x) &=& \frac{1}{8}\sum_{\langle \tilde{x},x\rangle} \sigma(\tilde{x})
\eeqa

while the model with a $U(1)$ chiral symmetry requires two real auxiliary fields $\sigma(\tilde{x})$ and $\pi(\tilde{x})$, such that

\beqa
S_{AF}[\sigma,\pi]   & =& \frac{N}{4g^2}\sum_{\tilde{x}}\left(\sigma^2(\tilde{x})+\pi^2(\tilde{x})\right ),
\\
\phib(x) &=& \frac{1}{8}\sum_{\langle \tilde{x},x\rangle} \Big(\sigma(\tilde{x}) + i\varepsilon(x)\pi(\tilde{x})\Big),
\eeqa

where $\varepsilon(x)$ is the parity of a lattice site ($1$ on even sites and $-1$ on odd sites). In the above expressions, the set of nearest dual sites $\tilde{x}$ surrounding the fixed lattice site $x$ is denoted as $\langle \tilde{x},x\rangle$ (see Fig.~\ref{fig:sites}). In this work we only consider these two classes of models.

It is easy to verify that $S_{GN}$ is invariant under $U(N)$ flavor transformations. When $m=0$, additional chiral symmetries emerge. The $Z_2$ model is invariant under $\chi_i(x) \rightarrow \varepsilon(x)\chi_i(x),\ \chib_i(x) \rightarrow -\chib_i(x) \varepsilon(x),\ \sigma(\tilde{x}) \rightarrow -\sigma(\tilde{x})$
while the $U(1)$ model is invariant under the additional $U(1)$ chiral symmetry $\chi_i(x) \rightarrow \mathrm{e}^{i\varepsilon(x)\theta/2}\chi_i(x),\  \chib_i(x) \rightarrow \chib_i(x) \mathrm{e}^{i\varepsilon(x)\theta/2},\ 
\sigma(\tilde{x}) \rightarrow \sigma(\tilde{x})\cos\theta + \pi(\tilde{x}) \sin\theta, \   
\pi(\tilde{x}) \rightarrow \pi(\tilde{x})\cos\theta - \sigma(\tilde{x}) \sin\theta$. The models contain a quantum critical point (QCP) separating a chirally symmetric phase (at small couplings) from a phase where the chiral symmetry is spontaneously broken (at large couplings). The symmetries that govern the QCP  needs proper analysis due to fermion doubling. Without such an analysis it is difficult to establish the continuum field theory that emerges at the critical point \cite{Chandrasekharan:2011mn}.

In the traditional MC approach, one integrates over the Grassmann fields and writes the partition function of the GN models as

\beqa
Z_{Z_2} &=& \int [\mathcal{D} \sigma]~e^{-S_{AF}[\sigma]}~ \Bigg\{\mbox{Det} D([\phib]) \Bigg\}^N,
\label{eq:pfz2}
\\
Z_{U(1)} &=& \int [\mathcal{D} \sigma \mathcal{D} \pi]~e^{-S_{AF}[\sigma,\pi]}~\Bigg\{\mbox{Det} D([\phib]) \Bigg\}^N,
\label{eq:pfu1}
\eeqa

In order to design a MC method the determinant terms in the above expressions have to be real and positive. In the $Z_2$ model since $\phib$ is real, the matrix elements of $D[\phib]$ are real. Hence, the determinant is real but not necessarily positive. In the case of the $U(1)$ model, $\phib$ is complex and so the matrix elements of $D[\phib]$  and its determinant can be complex. Hence, the $Z_2$ model as formulated in Eq.~(\ref{eq:pfz2}) suffers from a sign problem for all odd values of $N$, while the $U(1)$ model as  formulated through Eq.~(\ref{eq:pfu1}) suffers from a sign problem for all values of $N$.

\begin{figure*}[t]
\centering
\begin{tabular}{cc}
\includegraphics[width=2.0in]{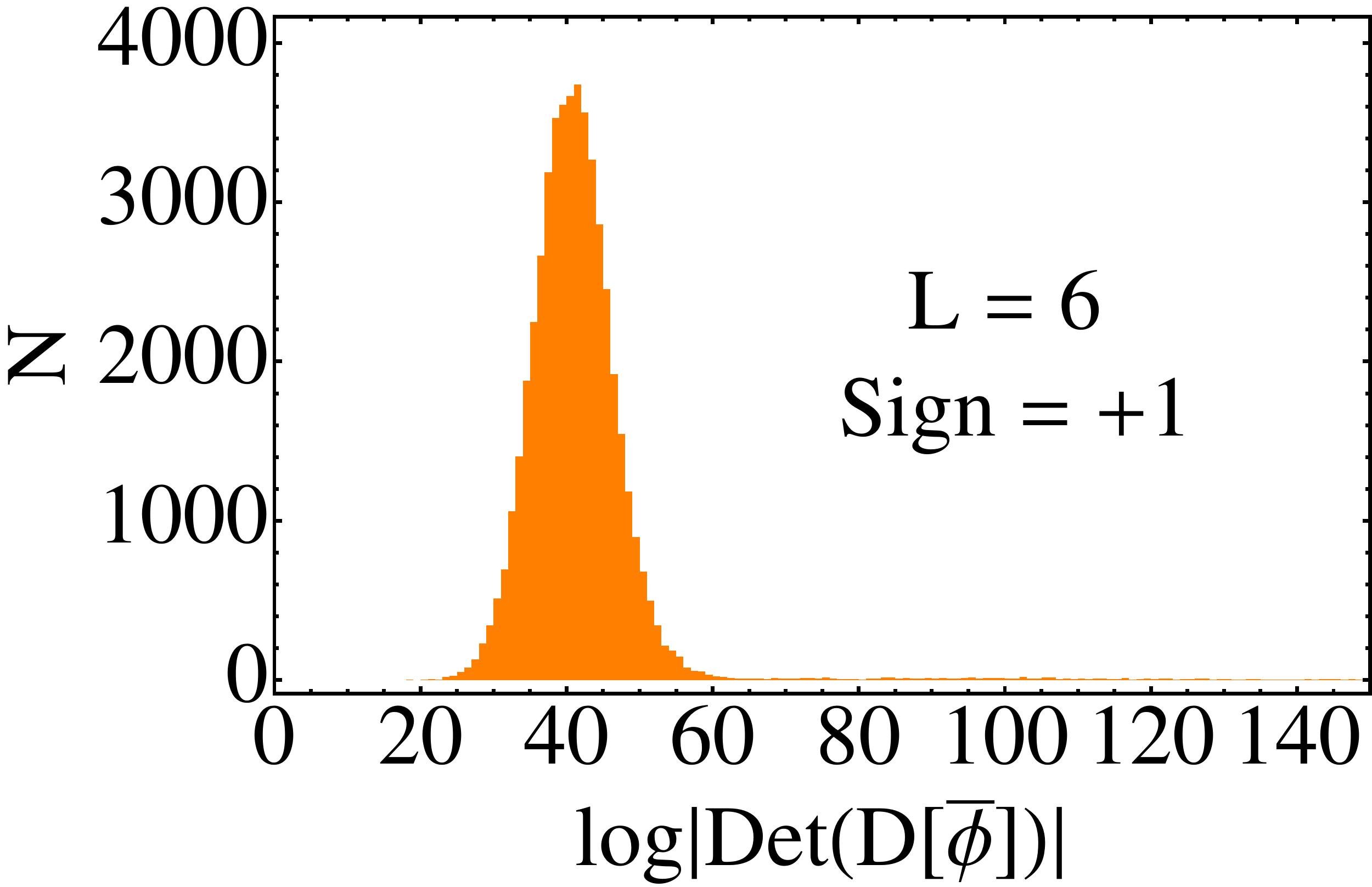} &
%\hskip0.1in
\includegraphics[width=2.0in]{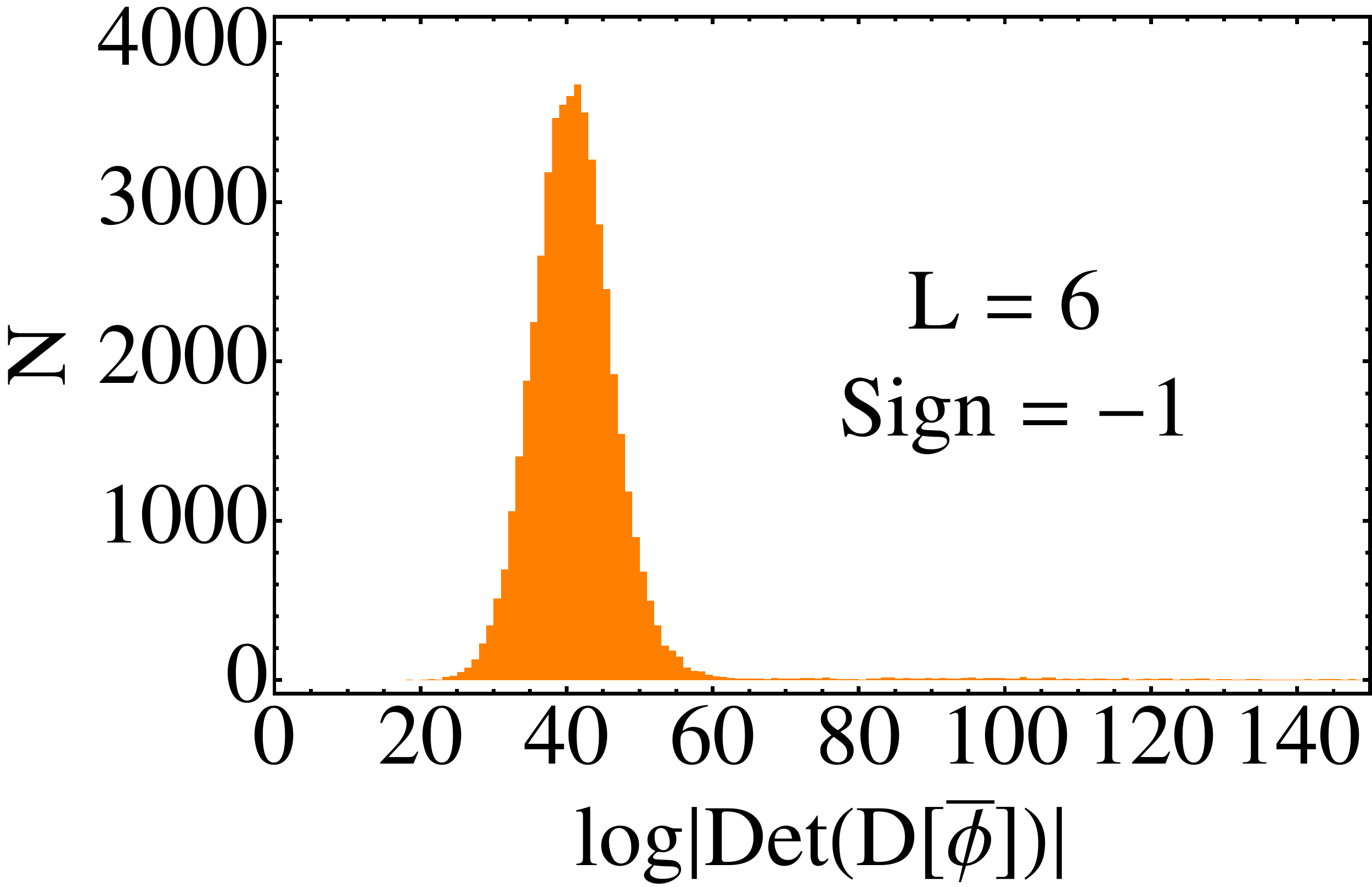} \\
%\hskip0.1in
\includegraphics[width=2.0in]{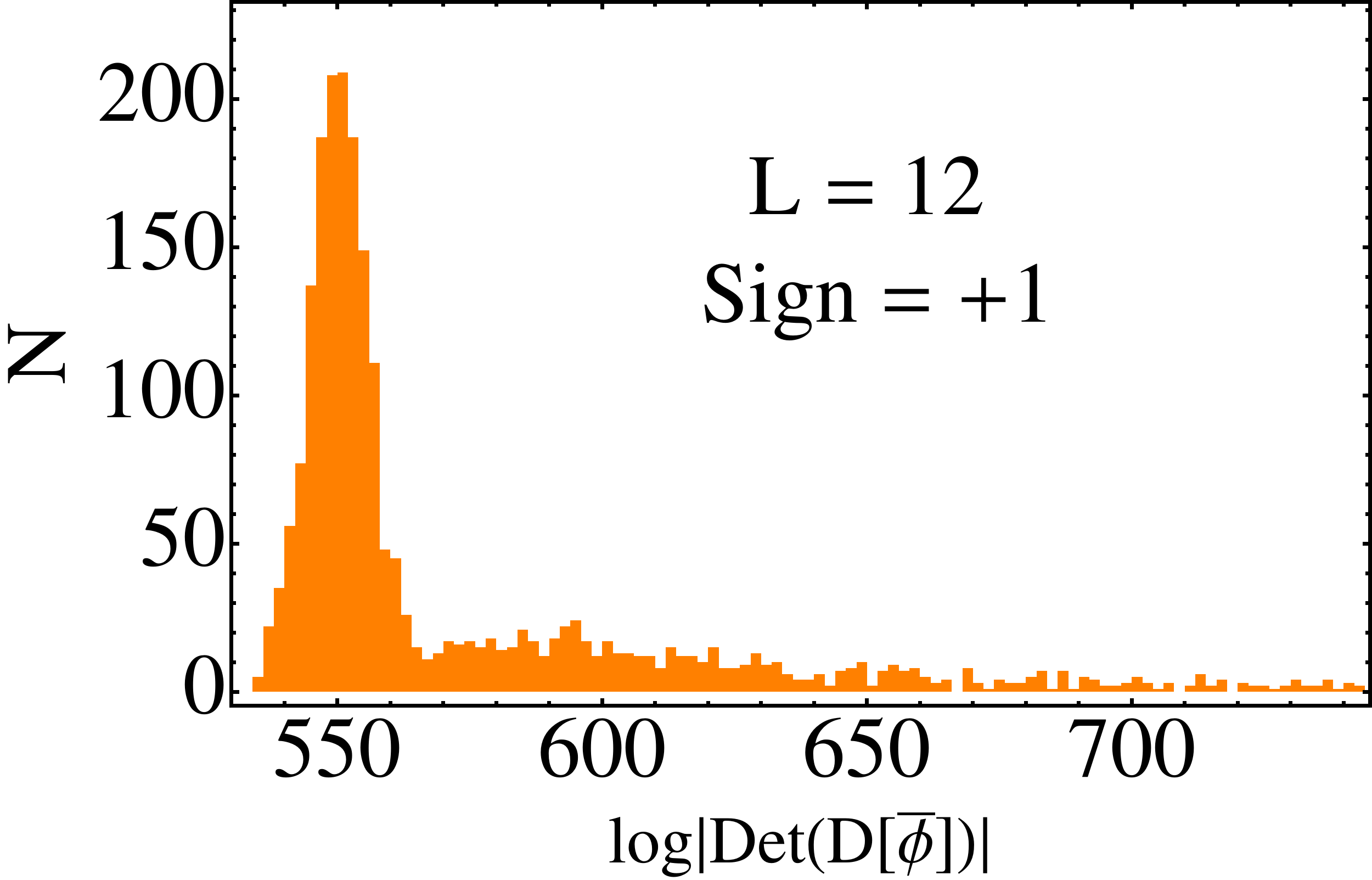} &
%\hskip0.1in
\includegraphics[width=2.0in]{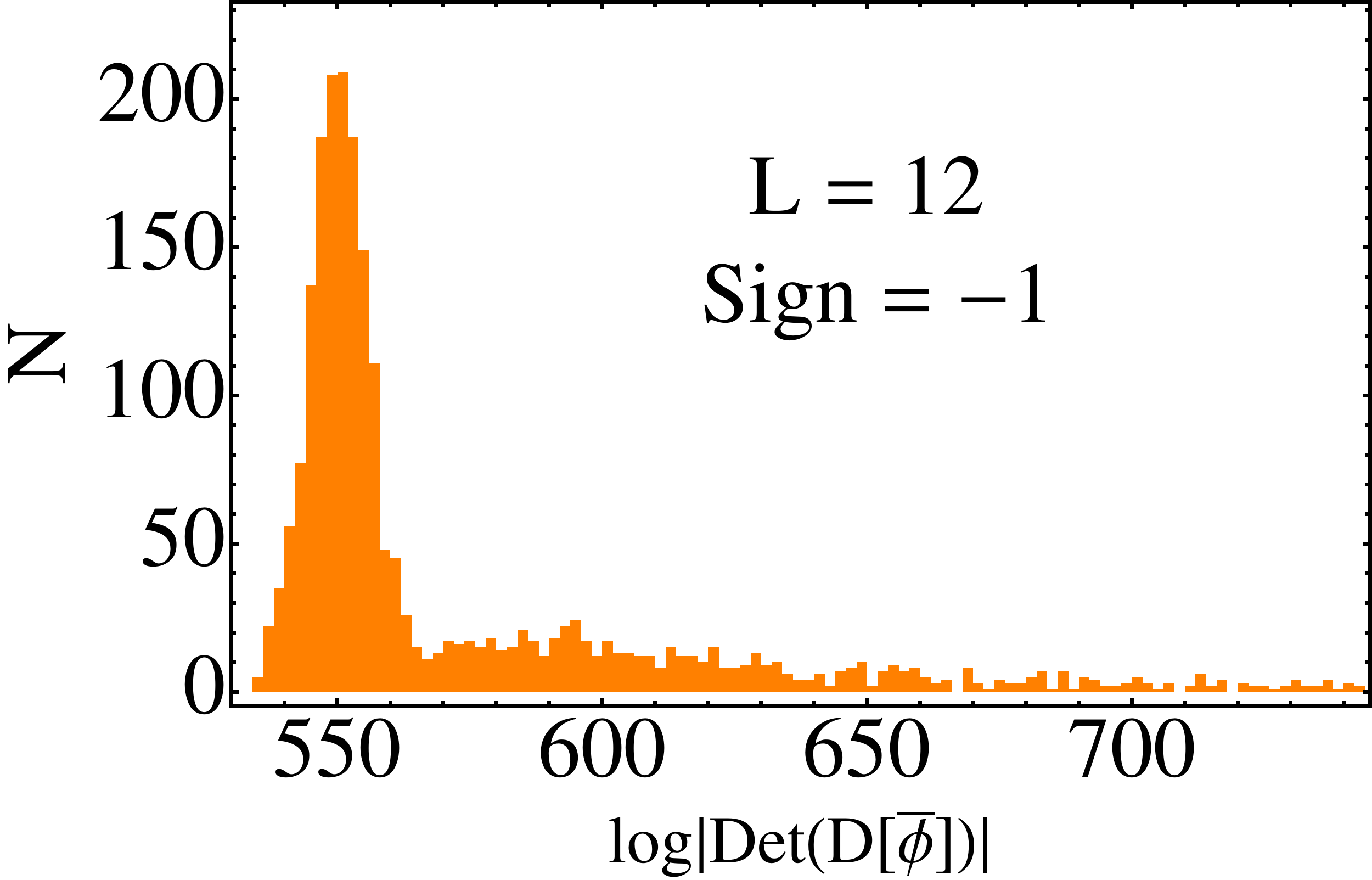}
\end{tabular}

\caption{Distributions of positive (left graphs) and negative (right graphs) weight configurations as a function of $\log|\mathrm{Det}(D[\phib])|$. One million configurations and 5000 configurations were generated at $6^3$ and $12^3$ lattices respectively. The distribution of positive configurations is almost identical to the one with negative configurations suggesting a severe sign problem.}
\label{fig:sign}
\end{figure*}

\section{Severity of the sign problem}

Earlier calculations in the $Z_2$ and $U(1)$ GN models have all been performed in the auxiliary field approach \cite{Hands:1995jq,Hands:2001aq,Christofi:2006zt}. The essential focus has been to understand the quantum phase transition and compute the critical exponents. These calculations have circumvented the sign problem by studying even $N$ in the $Z_2$ case or by introducing conjugate fermions with an opposite chiral charge in the $U(1)$. Inclusion of conjugate fermions changes the partition function from Eq.~(\ref{eq:pfu1}) to
\beq
Z^c_{U(1)} = \int [\mathcal{D} \sigma] [\mathcal{D} \pi]~e^{-S_{AF}}~
\Bigg|\mbox{Det} D([\phi]) \Bigg|^{2N},
\eeq
and changes the flavor symmetries to $U(N) \times U(N)$ while the chiral symmetry remains unchanged.

The $N=1$ model with $Z_2$ chiral symmetry was also studied in the auxiliary field approach using the HMC algorithm \cite{Karkkainen:1993ef}. Strangely, in this study the sign of the determinant was never discussed and seems to have been ignored. Since the results of the quantum critical behavior were in quantitatively agreement with large $N$ results (improved with Pad\'{e}-approximations), it may have been assumed that the sign problem was mild. If this is indeed true then statistically, positive sign configurations should dominate over negative sign configurations. The $Z_2$ model studied in \cite{Karkkainen:1993ef} is slightly different from the model studied here. The auxiliary fields $\sigma(x)$ also live on the main lattice site and the field $\phib$ appearing in the Dirac operator $D[\bar{\phi}]$ of Eq.~(\ref{eq:Dirac}), is defined as $\bar{\phi}(x) = \frac{1}{6} \sum_{\langle z,x\rangle} \sigma(z)$, where now $\langle z,x\rangle$ refers to the six nearest neighbor sites $z$ for a given site $x$. In order to study the sign problem, we generated several Gaussian random auxiliary field configurations according to the distribution
\beq
P(\sigma(x)) = \exp\Bigg(-\sum_{x}\Big\{ \sigma^2(x) - \frac{1}{2}\log(\pi)\Big\}\Bigg)
\eeq
and computed $\mathrm{Det}(D[\phib])$ for each of these configurations. We then separated the configurations into those with a positive determinant and those with a negative determinant. In Fig.~\ref{fig:sign} we plot the distribution of configurations with positive and negative determinants as a function of $\log|\mathrm{Det}(D[\phib])|$ for $6^3$ and $12^3$ lattices. As can be seen, the distribution of configurations with positive and negative weights are almost identical suggesting a severe sign problem rather than a mild one! Although we are not performing important sampling, our results clearly show that the sign problem must be studied carefully.

An important question to study is whether the HMC algorithm is getting trapped in the sector of configurations with positive weights (or negative weights). Note that, in the $Z_2$ models the only way to move from a positive weight sector to the negative weight sector is to pass through configurations which have almost zero weight assuming the step size in the HMC algorithm is small. Perhaps the suppression of the tunneling between the two sectors leads to long auto-correlation times or even lack of ergodicity. This argument also applies to $Z_2$ models with even $N$ \footnote{Simon Hands, private communication}. 

\section{Fermion bag approach}

We will now show that the sign problems in both the $Z_2$ and the $U(1)$ models discussed in section \ref{afa}, disappear in the fermion bag approach. The proof relies on the fact that any $k_i$-point correlation function involving the $i^{\mathrm{th}}$ flavor of staggered fermions defined through
\beqa
&& C_i(x_{i_1},...,x_{i_{k_i}}) = \int [d\chib_i d\chi_i] \mathrm{e}^{-\sum_{x,y} \ \chib_i(x) \ D_{xy}\ \chi_i(y)} 
\chib_{i}(x_{i_1})\chi_{i}(x_{i_1})\ ...\ \chib_{i}(x_{i_{k_i}})\chi_{i}(x_{i_{k_i}})
\label{eq:corr}
\eeqa
is positive semi-definite. This is due to the special properties of the free staggered fermion matrix. Indeed, using the ideas developed in the fermion bag approach  \cite{Chandrasekharan:2011mn}, we can write
\begin{equation}
C_i(x_{i_1},..,x_{i_{k_i}}) = \mathrm{Det}(D)\ \mathrm{Det}(G[\{x\}_i]) = \mathrm{Det}(W[\{x\}_i])
\label{eq:fbag}
\end{equation}
where $G[\{x\}_i]$ is the $k_i \times k_i$ matrix of propagators between the $k_i$ sites in the set $\{x\}_i \equiv x_{i_p}, \ p=1,..,k_i$ whose matrix elements are $G_{x_p,x_q} = {D}^{-1}_{x_p,x_q}$ and the matrix $W[\{x\}_i]$ is a $(V-k_i) \times (V-k_i)$ matrix identical to the matrix $D$ except that the sites in the set $\{x\}_i$ are dropped from the matrix. All the determinants appearing in Eq.(\ref{eq:fbag}) can be shown to be positive (or zero). The simplest way to see this is to consider the matrix $W$. Since it is exactly the same as the staggered fermion matrix with some sites removed, its eigenvalues come in complex conjugate pairs of the form $m\pm i\lambda$. Unpaired eigenvalues are always $m$ and they too come in pairs when the lattice is bipartite. When $m=0$ then the determinant can be exactly zero. Thus, $C_i(x_{i_1},..,x_{i_{k_i}}) \geq 0$. We will use this property to prove the absence of a sign problem in the fermion bag approach.

\begin{figure}[h]
\begin{center}
\includegraphics[width=1.0in]{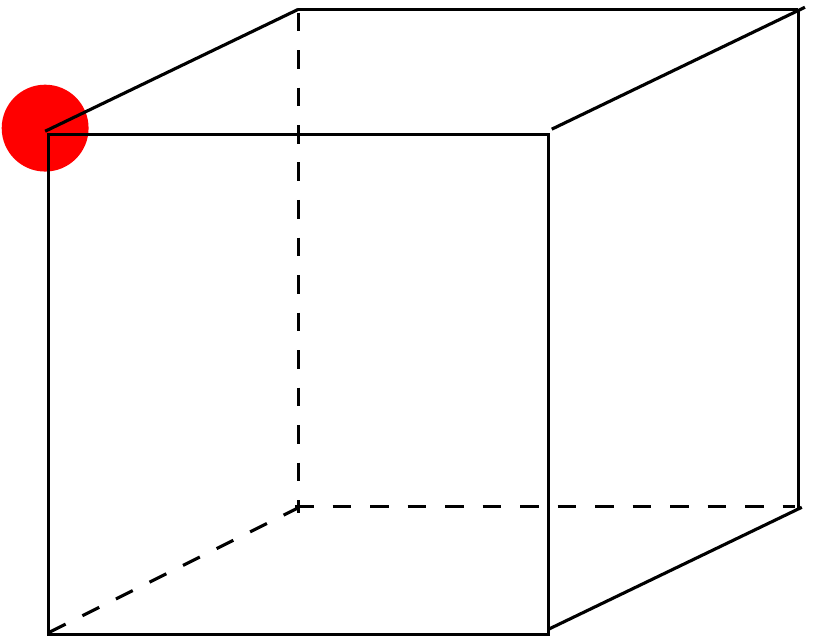}
\hskip0.1in
\includegraphics[width=1.0in]{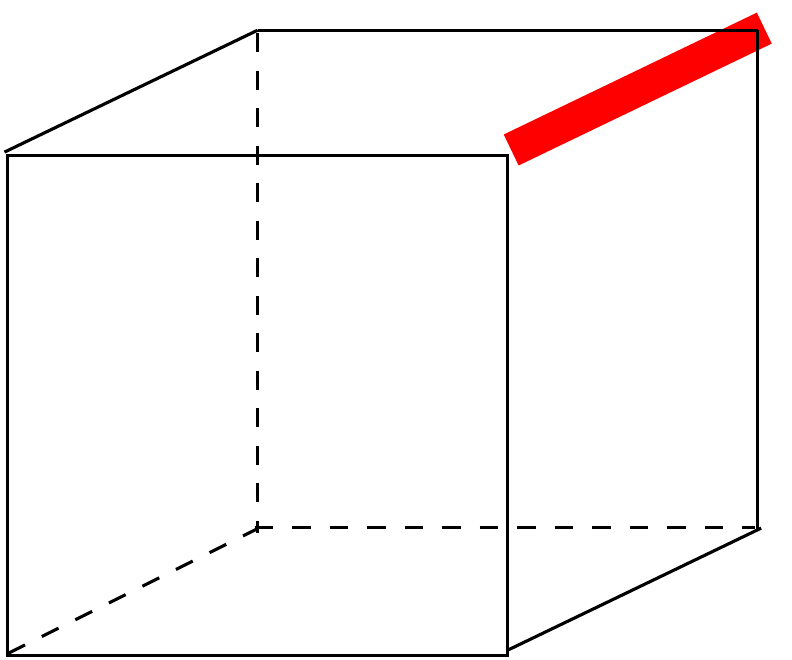}
\hskip0.1in
\includegraphics[width=1.0in]{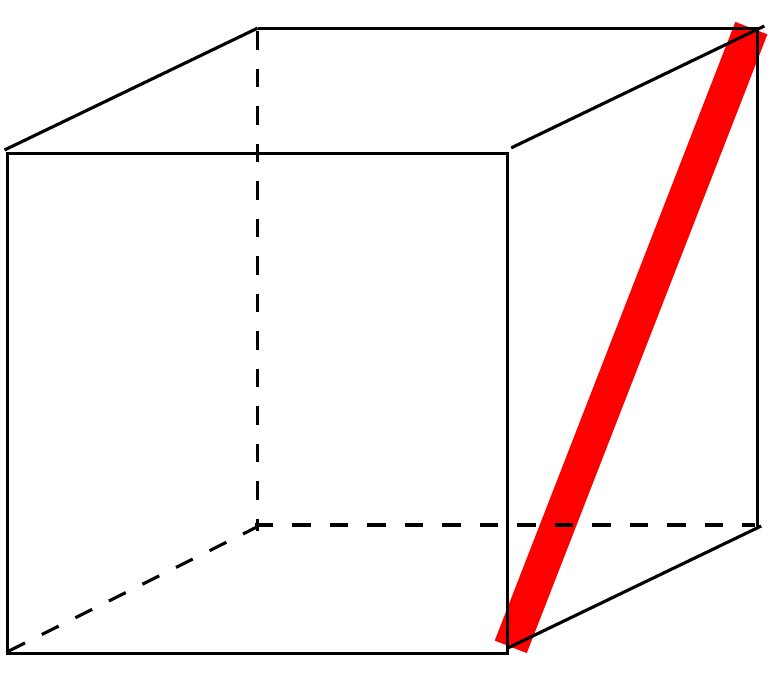}
\hskip0.1in
\includegraphics[width=1.0in]{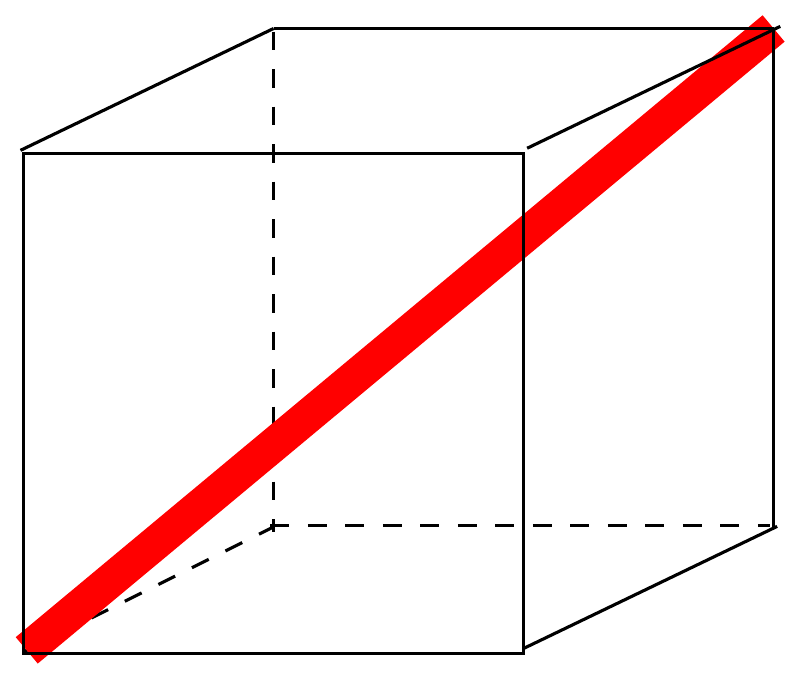}
\end{center}
\caption{An illustration of the four types of four-fermion couplings (or bonds) generated through the auxiliary field integration.  From left to right we have a $S$, $L$, $F$ and $B$ bond respectively.}
\label{fig:bonds}
\end{figure}

Instead of integrating out the fermion fields let us integrate out the auxiliary fields first and construct the appropriate four fermion action for the models. Let us first consider the $Z_2$ model. Each integral over the auxiliary field $\sigma(\tilde{x})$ on the dual site $\tilde{x}$ gives,
\beq
I_{\tilde{x}}  = \int d\sigma(\tilde{x}) \ \mathrm{e}^{-S_{AF} - \frac{\sigma(\tilde{x})}{8}\left(\sum_{ i, [x,\tilde{x}]} \chib_i(x)\chi_i(x)\right) }
  =  {\cal N}
\mathrm{e}^{ - S_I(\tilde{x}) },
\label{eq:Ix}
\eeq
where ${\cal N} = \sqrt{2\pi g^2/N}$ and 
\beqa
S_I(\tilde{x}) = -\frac{g^2}{128 N} \Big[\sum_{i,[x,\tilde{x}]} \chib_i(x) \chi_i(x) \Big]^2,\ \ 
\eeqa
is the effective four-fermion interaction term at each dual site $\tilde{x}$. The symbol $[x,\tilde{x}]$ denotes the set of all lattice sites surrounding the dual site $\tilde{x}$ (see Fig.~\ref{fig:sites}) . Thus, each integral generates many four-fermion couplings of the form $\chib_i(x)\chi_i(x) \chib_j(y) \chi_j(y)$ where $i$ and $j$ are arbitrary flavor indices and $x$ and $y$ are corners of the cube surrounding the dual site $\tilde{x}$. We can classify the possible couplings into four types based on the bonds $\langle xy\rangle$ connecting the corners $x$ and $y$. If the two corners are the same we refer to it as a site-bond or a $S$-bond. If the two corners are the two neighboring sites we get a $L$-bond (or a link-bond). Similarly, if the two corners are across a face diagonal or a body diagonal, we call the bonds $F$-bond and $B$-bond respectively. These four bond types are illustrated Fig.~\ref{fig:bonds}. 

Integration over all the auxiliary field variables yields the four-fermion interaction term of the action $S_{Z_2,{\rm int}} = \sum_{\tilde{x}} S_I(\tilde{x})$. Collecting the terms in each of the four types of four fermion couplings separately we see that
\beq
S_{Z_2,{\rm int}} = U_S \mathcal{B}_{S}  + U_L \mathcal{B}_{L} +  U_F \mathcal{B}_{F} + U_B \mathcal{B}_{B}
\eeq
where $U_S/4 = U_L/4 = U_F/2 = U_B = g^2/(64N)$ and 
\beq
\mathcal{B}_{\textrm{bond}} = \sum_{i,j, \langle xy\rangle \in {\rm bond}} \chib_i(x)\chi_i(x) \chib_j(y) \chi_j(y).
\eeq
Based on the above results, the partition function of the $Z_2$ model can be rewritten as
\beq
Z_{Z_2} = \int \prod_i [d\chib_i d\chi_i] \ \mathrm{e}^{-S_{Z_2}}.
\eeq
where $S_{Z_2}   \ =\  S_0 + S_{Z_2,{\rm int}}$ is the equivalent four-fermion action of the model. Here $S_0 = \sum_{x,y,i}\chib_i(x) D_{x,y} \chi_i(y)$ is the free fermion action.

In the fermion bag approach, each four-fermion coupling is represented as a bond and expanded in powers of the coupling. For example the four-fermion coupling of the type $\chib_i(x_p)\chi_i(x_p) \chib_j(x_q) \chi_j(x_q)$ can be denoted by the bond variable $b_{ij}(x_{p},x_{q}) = 0,1$, such that if it is $0$ then no bond is assumed to exist between the sites $x_p$ and $x_q$, otherwise the specific four-fermion coupling is inserted in the partition function. Due to the Grassmann nature of the couplings higher powers of the couplings do not exist. More details can be found in \cite{Chandrasekharan:2009wc}. Thus, in the fermion bag formulation, the partition function can be written as a sum over these bond configurations $[b]$, such that
\beqa
Z_{Z_2}&=& \sum_{[b]} U_S^{n_S} U_L^{n_L} U_F^{n_F} U_B^{n_B} 
\int \ \prod_i [d\chib_i d\chi_i] \ \mathrm{e}^{-S_0} \prod_{i} \chib_i(x_{i_1}) \chi_i(x_{i_2})...\chib_i(x_{i_{k_i}})\chi_i(x_{i_{k_i}})
\nonumber \\
&=& \sum_{[b]} U_S^{n_S} U_L^{n_L} U_F^{n_F} U_B^{n_B} \Big\{\prod_i C_i(x_{i_1},..,x_{i_{k_i}})\Big\}
\eeqa
where $n_S,n_L,n_F$ and $n_B$ are the total number of bonds of each type and the correlation function $C_i(x_{i_1},..,x_{i_{k_i}})$ was defined in Eq.(\ref{eq:corr}). A given bond configuration $[b]$ uniquely determines the $k_i$ sites $x_{i_1}....x_{i_{k_i}}$ (ordered in a consistent way). Since we argued above that $C_i(x_{i_1},..,x_{i_{k_i}}) \geq 0$ there is no sign problem in this expansion of the partition function for all non-negative values of $U_S$,$U_L$,$U_F$, $U_B$, any positive integer $N$ and real mass $m$.

In the case of the $U(1)$ model, we need to integrate over both the auxiliary fields $\sigma(\tilde{x}), \pi(\tilde{x})$ on every dual site. It is straightforward to verify that 
\beqa
I_{\tilde{x}}  &=& \int [d\sigma(\tilde{x}) d\pi(\tilde{x})] \ \mathrm{e}^{-S_{AF} - \frac{\sigma(\tilde{x})}{8}
\left(\sum_{ i, [ \tilde{x},x]} \chib_i(x)\chi_i(x)\right)}
\nonumber \\
&& \times \ \mathrm{e}^{ -i\frac{\pi(\tilde{x})}{8}
\left(\sum_{ i, [ \tilde{x},x]} \varepsilon(x)\chib_i(x)\chi_i(x)\right) } = {\cal N} \mathrm{e}^{-S_I(\tilde{x})}
\eeqa
where ${\cal N} =(4\pi g^2/N)$ and
\beqa
S_I(\tilde{x}) &=& \frac{g^2}{64 N} \Bigg\{
\Big[\sum_{i,[x,\tilde{x}]} \chib_i(x) \chi_i(x) \Big]^2 - \Big[\sum_{i,[x,\tilde{x}]} \varepsilon(x)\chib_i(x) \chi_i(x) \Big]^2\Bigg\},
\eeqa
Interestingly, the four-fermion couplings of the type $S$ and $F$ get canceled between the two terms in the above equation. On the other hand couplings of the type $L$ and $B$ survive so that the four-fermion action for the $U(1)$ model turns out to be
\beq
S_{U(1)} = S_0 + U_L \mathcal{B}_{L} + U_B \mathcal{B}_{B}
\label{eq:u1_action}
\eeq
with $U_L/4 = U_B = g^2/(16 N)$. Thus, the only difference between the $Z_2$ and $U(1)$ models is that the couplings $U_S = U_F = 0$ in the $U(1)$ model. Indeed these couplings break the $U(1)$ symmetry to a $Z_2$ symmetry as can be easily verified.
Since we already proved that the sign problem in the $Z_2$ model was absent for all non-negative values of $U_S$, $U_L$, $U_F$, $U_B$ and $N$ in the fermion bag formulation, the same is true for the $U(1)$ model as well.

\section{Results for $U(1)$ model}
It has been pointed out in~\cite{PhysRevD.51.5816} that the Thirring model is different from the Gross-Nevue(GN) model in many ways. The critical exponents in a lattice GN model with a $U_f(1) \times Z_2$ symmetry with staggered fermions have been computed and it was found that $\nu = 1.00(4)$ and $\eta = 0.754(8)$~\cite{Karkkainen:1993ef}. In a $U(1)$ lattice GN model simulation~\cite{Christofi:2006zt}, $\eta=0.904(50)$ has been found. However, those simulations are all based on the conventional Monte Carlo methods with auxiliary fields and an extra scale $m$ in the action. It could be tricky to carry out the chiral extrapolation analysis while simulations are closed to $U_c$. The most importantly, as we have pointed out in the previous sections, the original theory which has $U(1)$ flavor symmetry suffers from the sign problem, the conventional methods instead studied the modified theory which has $U(1) \times U(1)$ flavor symmetry with unchanged chiral symmetry. The expanding of the flavor symmetry can potentially affect the critical exponents and turn the model into different universality class. It is desirable to study the original model using a complete new method. 

In our previous paper~\cite{PhysRevLett.108.140404}, we used fermion bag approach to study the quantum critical exponents of the Thirring model that has the action
\beq
S_{Thirring} = S_0 + U_L \mathcal{B}_{L} 
\eeq
It has been shown that our results agree with the early studies by the conventional approaches, but our data are more accurate. By formulating the GN model in the fermion bag approach, in fact we can show that on the lattice the difference between the two models is an additional four-fermion coupling along the body diagonal after integrating out the auxiliary fields. It has the action in Eq.~(\ref{eq:u1_action}) with the extra coupling  $U_B = U_L/4$, and the original theory can be simulated without sign problem under the fermion bag approach. The simulation of $U(1)$ GN model becomes straight forward, we should be able to compare the Thirring model to GN model at the transition with fairly small amount of modifications of the previous simulation. Before we start our simulations, as one can see, the body diagonal coupling is just quarter of the link coupling, it also preserves the $U(1)$ chiral symmetry. We expect this extra term should not change the universality class of  the GN model compared to the Thirring model, both models can have the same critical exponents. In the simulations, we actually let $U_L=U_B=U$, we believe this little modification should not have the affect on the critical exponents. 

\begin{figure*}[t]
\begin{center}
\hbox{
\includegraphics[width=2.0in]{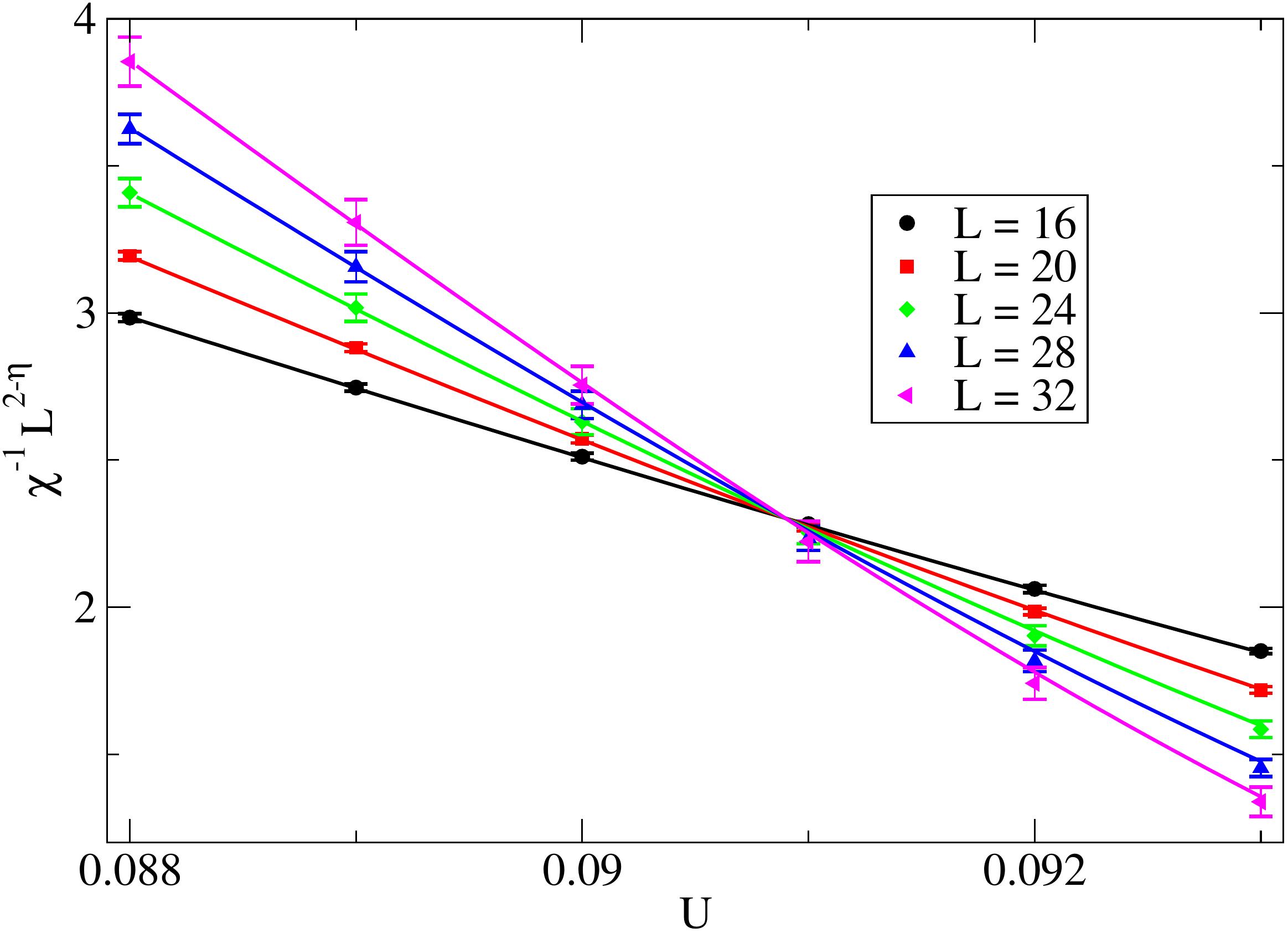}
\includegraphics[width=2.0in]{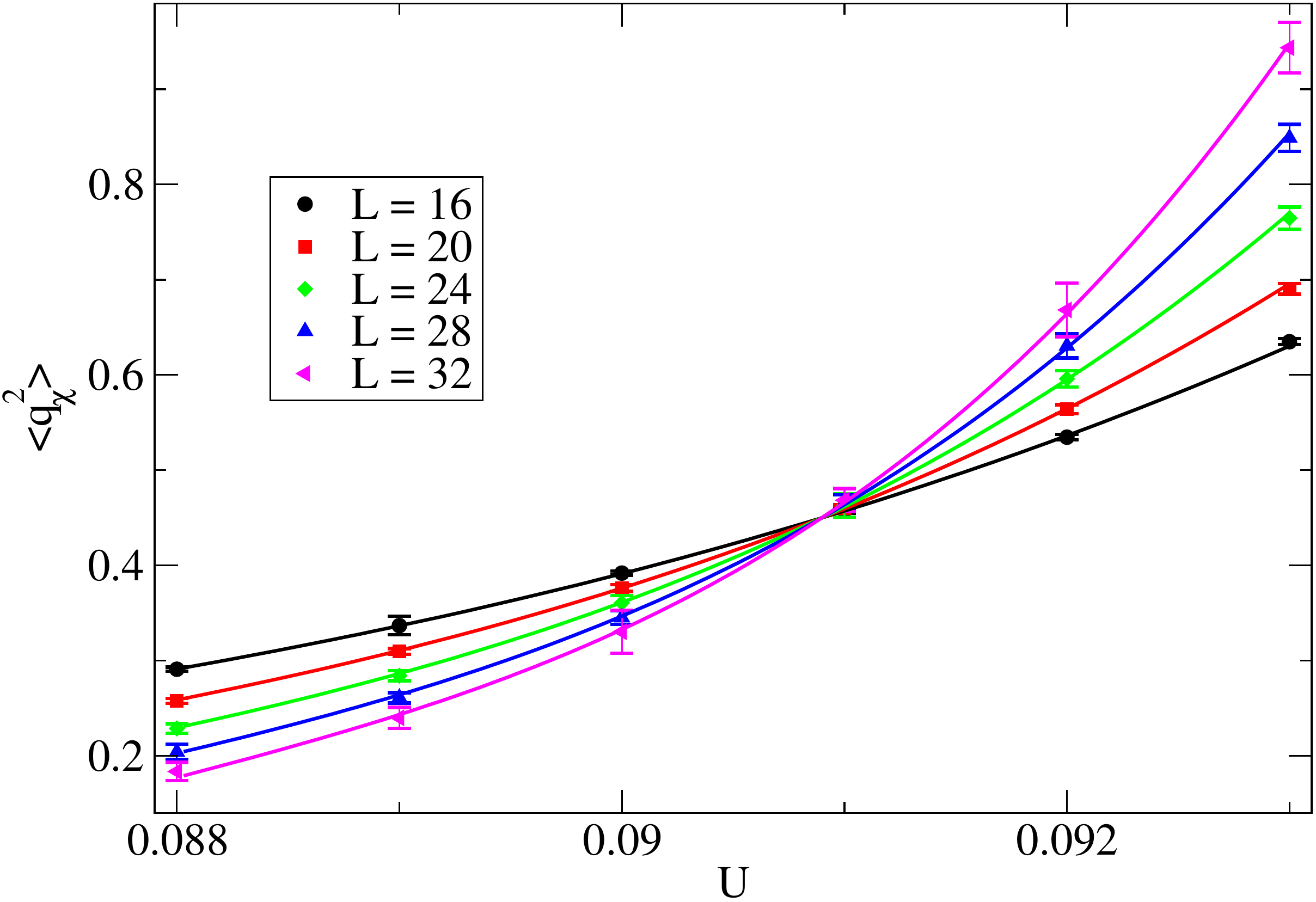}
\includegraphics[width=2.0in]{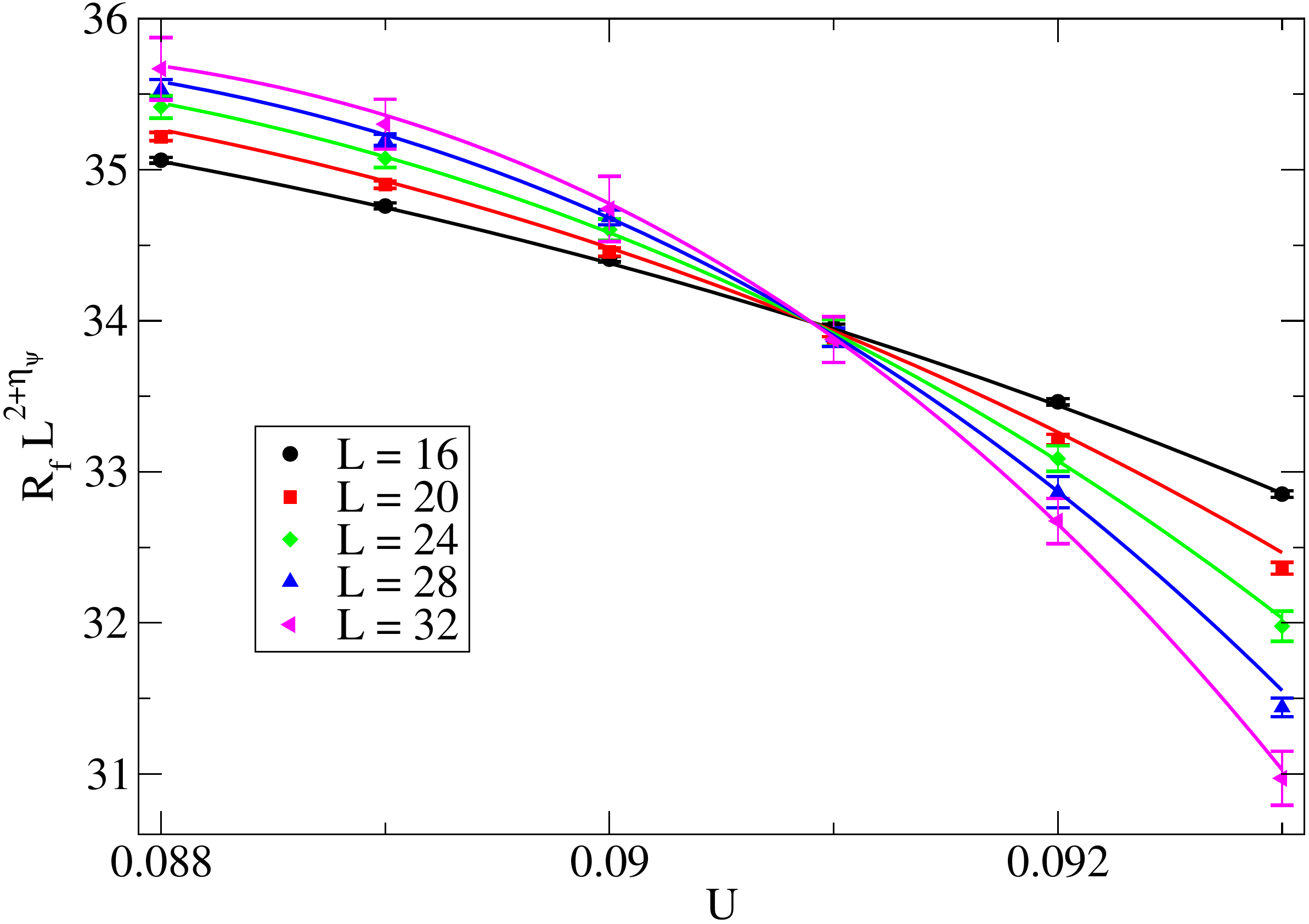}
}
\caption{Plots of $\chi^{-1} L^{2-\eta}$, $\langle q_\chi^2\rangle$ and $R_f L^{2+\eta_\psi}$ as a function of $U$ for $L$ from $16$ to $40$. The solid lines show the combined fit which give $U_c = 0.0909(1) , \nu = 0.88(1), \eta=0.63(1)$ and $\eta_\psi=0.37(1)$ with $\chi^2/d.o.f = 0.89$}
\label{fig:scaling}
\end{center}
\end{figure*}

In order to study the critical exponent, we focused on three observables (Let $L$ be the lattice size): The chiral condensate susceptibility 
\beq
\chi = \frac{1}{2 L^3}\sum_{x,y}\langle\psib_x\psi_x\psib_y\psi_y\rangle,
\eeq
the chiral winding number susceptibility 
\beq
\langle q^2_\chi \rangle =\langle \frac{1}{3}\sum_\alpha(q^2_\chi)_\alpha \rangle,
\eeq , and the ratio of fermion two-point correlator
\beq
R_f=C_F(L/2-1)/C_F(1), 
\eeq
The details of those observables are explained in \cite{Chandrasekharan:2011vy}. Since the fermions are exactly massless, in the vicinity of $U_c$ we expect these three observables to satisfy the following simple finite size scaling relations:

\beqa
\label{eq:scaling}
\chi^{-1} L^{2-\eta} = \sum_{k=0}^3 f_k \left[(U-U_c) L^{\frac{1}{\nu}}\right]^k \nonumber\\
\langle q_\chi^2 \rangle = \sum_{k=0}^3 \kappa_k \left[(U-U_c) L^{\frac{1}{\nu}}\right]^k \nonumber\\
R_f L^{2+\eta_\psi} = \sum_{k=0}^3 p_k \left[(U-U_c) L^{\frac{1}{\nu}}\right]^k 
\eeqa
where we have kept the first four terms in the Taylor series of the corresponding analytic functions. Our goal is to compute the critical exponents $\eta$, $\nu$ and $\eta_\psi$ at the quantum critical point. Plots of our data is shown in Fig.~\ref{fig:scaling}.  Since our data fits very well to the expected scaling form for a whole range of lattice sizes, we feel confident that the corrections to scaling are small. 

We compared the critical exponents from $U(1)$ GN model to the Thirring model in Table~\ref{table:comparision}. There are slight deviations between two models. However, by combining data sets from two models, we can fit the same scaling relations quite well with $\chi^2/d.o.f = 1.19$. The joined fitting results are listed as ``Combined'' in the table. It is clear that those critical exponents are completely different from mean-field analysis and early MC studies by conventional methods~\cite{Christofi:2006zt}. Our data shows that $U(1)$ GN model and Thirring model belong to the same universality class. It will be very interesting to carry out the same study on the $Z_2$ GN model and compare to the early results~\cite{Karkkainen:1993ef}. Since the $U(1)$ chiral symmetry has been broken explicitly by the face diagonal bond, the critical behaviors can be different from the $U(1)$ model.

\begin{table}[h]
\begin{center}
\begin{tabular}{c | c c c c}
\hline\hline
 & $\nu$ &$\eta$& $\eta_\psi$ & $\chi^2/d.o.f.$  \\
 \hline
 Thirring & 0.85(1) & 0.65(1) & 0.37(1) &1.3 \\
\\
$U(1)$ GN & 0.88(1)&0.63(1)&0.37(1)& 0.89\\
\\
Combined & 0.85(1) & 0.64(1) & 0.37(1) & 1.19\\
\hline\hline
\end{tabular} 
\caption{Results from the fit of the data to Eqs.~(\ref{eq:scaling}). In addition to fit Thirring and $U(1)$ model separately,  we also combined two data sets and fit all the critical exponents using the same scaling relations.}
\label{table:comparision}
\end{center}
\end{table}

\section{Conclusions}

The fermion bag approach provides an alternative approach to fermion field theories where solutions to new sign problems emerge naturally. Here we have demonstrated that some sign problems in the auxiliary field formulation of GN models, especially with $Z_2$ and $U(1)$ chiral symmetries, disappear in the fermion bag approach. By using the fermion bag approach, we have also shown the $U(1)$ GN model and Thirring model belong to the same universality classes. While we have not shown here, we can solve sign problems in some lattice field theories containing both dynamical boson and fermion fields with similar chiral symmetries. In these more complex models, the solutions emerge when bosons are formulated in the world-line approach and the fermions are formulated in the bag approach. Such an approach to quantum field theories was proposed in \cite{Chandrasekharan:2008gp}.

Sign problems in other fermion models with more complex symmetries are also solvable in the fermion bag approach. However, in many interesting cases the Boltzmann weight of a fermion bag, although non-negative, turns out to be a fermionant instead of a determinant \cite{Chandrasekharan:2011an}. Since the computation of the fermionant can be exponentially hard, the fermion bag approach loses its practical appeal in such cases. Still, we believe that there are many other interesting models where the weight of the fermion bag continues to be positive and computable with polynomial effort.

\section*{Acknowledgments}
I would like to thank my collaborator S. Chandrasekharan for the wonderful collaboration and fruitful discussions. This work was supported in part by the Department of Energy grants DE-FG02-05ER41368 and DE-FG02-00ER41132.

\section*{References}
\bibliographystyle{iopart-num}
\bibliography{ref}

\end{document}